\documentclass[aps,prd,twocolumn,eqsecnum]{revtex4}
\begin{document}

\title{Search for  combinations of thermal n-point
functions with analytic extensions}

\author{H. Arthur Weldon}
\affiliation{Department of Physics, West Virginia University, Morgantown WV
26506-6315}

\date{\today}

\begin{abstract}
The $2^{n}$ different n-point functions that occur in real-time thermal field theory
are Fourier transformed to real energies. Because of branch cuts
in various energy variables, none of these functions can be extended
analytically to  complex energies.  The known  linear combinations that
form the  fully retarded and   advanced functions can be extended 
analytically. It is proven that no other linear combinations have an
analytic extension to complex energies. 
\end{abstract}

\pacs{11.10.Wx, 03.65.Db, 02.30.Fn}

\maketitle

\section{INTRODUCTION}

\subsection{Analyticity at $T=0$}

In zero-temperature field theory the analytic properties of 
amplitudes is of fundamental importance. Physical scattering amplitudes
are functions of real energies but they can be extended analytically  into certain
regions of the complex energy planes. As a consequence, scattering
amplitudes are the boundary values of analytic functions when the
energy arguments are real.
 
Crossing symmetry at zero temperature  is a property of amplitudes for
different processes that are analytic continuations of each other.
Without analyticity there could be no crossing symmetry. In the
simplest case of $2\to 2$ scattering for equal mass particles, the
amplitude is a function of the Lorentz invariant variables $s,t,u$
subject to the constraint $s+t+u=4m^{2}$.  In the direct channel
$s>4m^{2}$ while $t$ and $u$ are negative. In one of the crossed
channels $s$ and $u$ are negative but $t>4m^{2}$. Two properties of
the scattering amplitude are important: (1) Starting from real
energies in either  physical region, the amplitude can be extended
analytically into a certain  region of the complex energy plane. For
example, the direct channel amplitude can be extended analytically
into the upper half of the complex $s$ plane but not into the lower
half because of a branch cut. Similarly for the cross channel.  (2)
The location of the branch cuts is such that the direct channel
amplitude can be analytically continued to obtain the cross channel
amplitude. Consequently, one function of the complex energies, when
evaluated at the appropriate real energies, can yield the scattering
amplitude for three different physical reactions \cite{ELOP}. 

\subsection{$T\neq 0$ Background}

The real-time formulation of thermal field theory requires a
mixture of time-ordered and anti-time-ordered products that give
$2^{n}$ different n-point functions \cite{Klaus,Michel,Ashok}. The
Fourier transforms 
are functions of the external energy-momenta $p_{j}$:
\begin{equation}
G_{\alpha_{1}\alpha_{2}...\alpha_{n}}(p_{1},p_{2},...p_{n}).\label{G}
\end{equation}
Each suffix $\alpha_{j}$ has two possible values, 1 or 2.  These
functions arise naturally in the time-path formulation. [See Eq.
(\ref{contour}) below.]  However,  it is often
convenient to employ various linear combinations of these basic functions.
Keldysh introduced a different set \cite{Kel} that has been
used extensively by Heinz and collaborators \cite{Ulrich1,Ulrich2,Ulrich3,FG}. 
Aurenche and Becherawwy  used the
transformations  which simplify the two-point functions
and apply them to the n-point functions and introduced what is
known as the $R/A$ basis \cite{Patrick} that has important physical
applications \cite{Brandt1,Brandt2}. Although it is not Feynman's fault,
there is an
$F/\overline{F}$ basis that was motivated by zero-temperature Feynman
propagators.  The algebraic relations between these various choice
of bases was  clarified in two papers by van Weert et al
\cite{Chris1,Chris2}. The analytic properties of the amplitudes was not treated
in these investigations. 
 
\paragraph*{The question of analyticity.}
For each $n$ there are $n$ fully-retarded thermal n-point functions and
$n$ fully-advanced thermal n-point functions, as reviewed in Sec. IV.  All of
these are linear combinations  of the basic functions in Eq.
(\ref{G}) and all of these can be analytically extended to certain domains of
the complex energies $p_{j}^{0}$.  The simplicity of these results leads to the
question of whether there are other linear combinations of the basic functions
in (\ref{G}) that can be analytically extended to other domains of complex
energy.

One can sharpen the question by incorporating the results of Evans \cite{Tim}. 
He showed that the single  n-point
function of the imaginary-time formulation  can be
analytically extended away from the discrete, imaginary energies to
various domains of  complex energies. For $n=4$ there are 32  thermal 4-point
functions that have
analytic extensions; the fully retarded and advanced functions account for only 8
of these.  For
$n=5$ there are new 370 thermal 5-point functions that have analytic
extensions; the fully retarded and advanced functions account for only 10 of
these.   The question addressed here is whether any of
these new n-point functions that have analytic extensions can be expressed as
linear combination of the basic functions in Eq. (\ref{G}). 

\paragraph*{Outline.}
Section II treats the thermal two-point functions and shows why particular 
linear combinations do have analytic extensions. Section III computes the
Fourier transforms of the thermal n-point functions and shows why none of the
$2^{n}$ functions can be extended analytically to complex energy. 
Furthermore, none of these functions is the difference of two analytic
functions across a branch cut. Section IV summarizes known results about the
fully retarded and the fully advanced n-point functions.  In particular, the
retarded can be analytically extended to a domain in which  one $p_{j}^{0}$ has
a positive imaginary part and all the others have negative imaginary parts; the
advanced can be analytically extended to a domain in which  one $p_{j}^{0}$ has
a negative imaginary part and all the others have positive imaginary parts. 
Section V shows that any
$n$-point function that is analytic in a more general domain, i.e. one in which
at least two energies have positive imaginary parts and at least two energies
have negative imaginary parts,  cannot be a linear combination of the real-time
thermal
$n$-point functions.

\section{\lowercase{n}=2 Example}

It is instructive to examine  two-point functions
in preparation for the more general results
in Sec. III-V.  Throughout the paper, the thermal average of any operator
$\Omega$ will be denoted
\begin{displaymath}
\sum_{a}P_{a}\langle a|\Omega|a\rangle,
\end{displaymath}
where the summation is over a completes set of energy eigenstates  
$|a\rangle$  with energy
$E_{a}$. Here
$P_{a}$ is the thermal probability of state $|a\rangle $:
\begin{equation} P_{a}=e^{-\beta E_{a}}\Big/\text{Tr}[e^{-\beta H}],
\label{Pa}\end{equation}
and $\beta$ is the inverse temperature.

\subsection{$G_{11}$ propagator} The simplest 2-point
function is the thermal average of the time-ordered product
\begin{displaymath}
G_{11}(x_{1},x_{2})=-i\sum_{a}P_{a}
\langle a|T\big[\phi(x_{1})\phi(x_{2})\big]|a\rangle.
\end{displaymath}
The Fourier transform is
\begin{eqnarray}
\int d^{4}x_{1}d^{4}x_{2}
e^{i(p_{1}\cdot x_{1}+p_{2}\cdot x_{2})}
G_{11}(x_{1},x_{2})\nonumber\\
=2\pi\delta(p_{1}^{0}{+}p_{2}^{0})G_{11}(p_{1},p_{2}).
\nonumber\end{eqnarray}
Note that $G_{11}(p_{1},p_{2})$ is proportional to
$(2\pi)^{3}\delta^{3}(\mathbf{p}_{1}+\mathbf{p}_{2})$.
The Fourier transform of the space dependence will be expressed in
terms of  
\begin{equation}
\phi_{j}\equiv\int\! d^{3}x_{j}\, e^{-i\mathbf{p}_{j}\cdot\mathbf{x}_{j}}
\phi(0,\mathbf{x}_{j}),
\label{space}\end{equation}
which contains the three-momentum $\mathbf{p}_{j}$. 
The Fourier transform of the time dependence can be performed using the
Hamiltonian evolution:
\begin{equation}\phi(t_{j},{\bf
x}_{j})=e^{iHt_{j}}\phi(0,{\bf
x}_{j})e^{-iHt_{j}}.
\label{time}\end{equation} The dependence on the
energies $p_{j}^{0}$ comes  through the  retarded resolvents
\begin{equation}
{\cal G}_{j}^{R}={1\over E_{a}+p_{j}^{0}-H{+}i\epsilon}.\label{ret}
\end{equation}
Working this out easily gives
\begin{eqnarray}
G_{11}(p_{1},p_{2})\!=\!\sum_{a}\!P_{a}\Big[\langle
a|\phi_{1}{\cal G}_{1}^{R}\phi_{2}|a\rangle{+}\langle
a|\phi_{2}{\cal G}_{2}^{R}\phi_{1}|a\rangle\Big].\label{n=2}
\end{eqnarray} 
The spectrum of $H$ is positive, definite. For a fixed value of $E_{a}$,
the first term has a semi-infinite branch cut in $p_{1}^{0}$ that extends
from some finite energy to $\infty$. 
Because the range of $E_{a}$ is unbounded when summed over, the first term
has a branch cut running the entire length of the real axis, $-\infty
<p_{1}^{0}<\infty$. The first term can be extended analytically from real
$p_{1}^{0}$  to any complex value with
$\text{Im}(p_{1}^{0})>0$ without having to cross the branch cut.  The
second term has a branch cut for
$-\infty <p_{2}^{0}<\infty$ and can be extended analytically from real
$p_{2}^{0}$ to any complex value satisfying
$\text{Im}(p_{2}^{0})>0$. 
 Since  the complex energies must satisfy
$p_{1}^{0}{+}p_{2}^{0}=0$, these two regions do not overlap.
Hence there is no domain of complex energies for which the  time-ordered
thermal propagator is naturally analytic. The qualifying word `naturally' 
is included because the dynamics may allow one of the terms in Eq. (\ref{n=2}) to be
analytically continued through the branch cut. 

\subsection{$G_{12}$ propagator}  In real-time thermal field
theory there are four types of propagators, labeled $G_{11}$,
$G_{12}$, $G_{21}$, $G_{22}$ \cite{Klaus,Michel,Ashok}. The second of these is
\begin{equation}
G_{12} (x_{1},
x_{2}) ={-}i\sum_{a,b}C_{ab}\langle
a|\phi(x_{2})|b\rangle\langle b|\phi(x_{1})|a\rangle\nonumber
,\end{equation} where
\begin{equation}
C_{ab}
=e^{\sigma(E_{a}-E_{b})}P_{a},
\label{cab}\end{equation}
and $0\le \sigma\le \beta$.  
The Fourier transform is
\begin{displaymath}
G_{12}(p_{1},p_{2})=-ie^{-\sigma p^{0}_{2}}\sum_{a}P_{a}
\langle a|\phi_{2}\,2\pi\delta[E_{a}{+}p_{2}^{0}{-}H]\,
\phi_{1}|b\rangle.
\end{displaymath}
The Dirac delta function forces $p_{2}^{0}$ to be real and therefore
$p_{1}^{0}$ is real.  No analytic extension is possible.

\subsection{Linear combination with an analytic extension} 

Although neither $G_{11}$ nor $G_{12}$ can be extended analytically to
complex energy, there is a simple linear combination of the two that can
be extended. 
This requires the
advanced resolvent, 
\begin{equation}
{\cal G}_{j}^{A}={1\over E_{a}+p_{j}^{0}-H{-}i\epsilon}.\label{adv}
\end{equation}
which is related to the retarded resolvent by
\begin{displaymath}
{\cal G}_{j}^{R}+i2\pi\delta[E_{a}{+}p_{j}^{0}{-}H]
={\cal G}_{j}^{A}.\end{displaymath}
The following linear combination of $G_{11}$ and
$G_{12}$ changes the ${\cal G}_{2}^{R}$ in Eq. (\ref{n=2})
to ${\cal G}_{2}^{A}$: 
\begin{eqnarray}
&&G_{11}(p_{1},p_{2})-e^{\sigma p_{2}^{0}}G_{21}(p_{1},p_{2})\\
&&\hskip1cm =\sum_{a}\!P_{a}\Big[\langle
a|\phi_{1}{\cal G}_{1}^{R}\phi_{2}|a\rangle{+}\langle
a|\phi_{2}{\cal G}_{2}^{A}\phi_{1}|a\rangle\Big].\nonumber
\end{eqnarray}
The  term containing ${\cal G}_{1}^{R}$ can be extended analytically
to
$\text{Im}(p_{1}^{0})>0$; the term containing ${\cal G}_{2}^{A}$
can be analytically extended to
$\text{Im}(p_{2}^{0})<0$. Since $p_{1}^{0}{+}p_{2}^{0}=0$ these two conditions
are compatible. This combination is known as the retarded propagator:
\begin{equation}
R(\buildrel\circ\over{p_{1}},p_{2})=G_{11}(p_{1},p_{2})-e^{\sigma
p_{2}^{0}}G_{21}(p_{1},p_{2}),
\end{equation}
where the circumflex over $p_{1}$ will be used to indicate that
$p_{1}^{0}$ is the energy that can have a positive imaginary part. 
 
Starting with Eq. (\ref{n=2}) there is another way to produce a 2-point function
that can be analytically extended, viz. change ${\cal G}_{1}^{R}$ to 
${\cal G}_{1}^{A}$. This produces the advanced two-point function
$A(p_{1},\buildrel\circ\over{p_{2}})$
which is analytic for $\text{Im}(p_{1}^{0})<0$
 and $\text{Im}(p_{2}^{0})>0$.

\section{\lowercase{n}-point functions in the real-time formulation}

 Thermal n-point functions are defined
\cite{Klaus,Michel,Ashok}
  in the complex time plane
on a contour that consists of two parts: $C_{1}$ runs along the real time axis
from
$-\infty$ to $\infty$; $C_{2}$ runs anti-parallel to the real time axis
from $\infty-i\sigma$ to $-\infty-i\sigma$.  The parameter $\sigma$
lies in the range $0\le \sigma\le \beta$.  A scalar field $\phi(x)$ may be
defined on contours
$C_{1}$ and $C_{2}$ by introducing
\begin{equation}
\Phi_{\alpha}(x)=\left\{\begin{array}{cc} \phi(x) & \text{if}\;
\alpha=1\\
e^{\sigma H}\phi(x)e^{-\sigma H} & \text{if}\;
\alpha=2\end{array}\right. 
\end{equation}
The thermal Green functions are
\begin{eqnarray}
&&G_{\alpha_{1}\alpha_{2}...\alpha_{n}}(x_{1},x_{2},...x_{n})
\label{contour}\\
&&\hskip0.5cm=(-i)^{n{-}1}\sum_{a}P_{a}
\langle a\big|T_{C}\big[\Phi_{\alpha_{1}}(x_{1})...
\Phi_{\alpha_{n}}(x_{n})\big]\big|a\rangle ,\nonumber
\end{eqnarray}
where $T_{C}$ denotes contour ordering.
Complex conjugation of a Green
function interchanges  the indices, $1\leftrightarrow 2$,
\begin{eqnarray}
&&\big[G_{\alpha_{1}\alpha_{2}...\alpha_{n}}(x_{1},x_{2},...x_{n})
\big]^{*}\\
&&\hskip1cm=(-1)^{n{-}1}G_{\overline{\alpha}_{1}\overline{\alpha}_{2}...
\overline{\alpha}_{n}}(x_{1},x_{2},...x_{n})\nonumber,
\end{eqnarray}
where $\overline{1}=2$ and $\overline{2}=1$.

When all the fields are on contour $C_{1}$ the result is
the thermal average of the time-ordered product of n fields:
\begin{equation}G_{1...1}(x_{1},...x_{n})
=({-}i)^{n{-}1}\sum_{a}P_{a}\langle
a|T\big[\phi(x_{1})...\phi(x_{n})\bigr]|a\rangle.
\label{1111}\end{equation} When all the fields are on $C_{2}$ the result
is thermal average of the anti-time-ordered product: 
 \begin{equation}
G_{2...2} (x_{1},...
x_{n}) =({-}i)^{n{-}1}\sum_{a}P_{a}\langle
a|\widetilde{T}\bigl[\phi(x_{1})...\phi(x_{n})\bigr]|a\rangle,
\label{G2222x}\end{equation}
where $\tilde{T}$ denotes anti-time-ordering.

In the general   case,  $\ell$ fields  are on contour
$C_{2}$ and the remaining
$n{-}\ell$ are on contour $C_{1}$. For example, if
$x_{1},x_{2},...x_{\ell}$ are the coordinates of the fields on $C_{2}$
and  $x_{\ell{+}1},...x_{n}$ are the coordinates of the fields on
$C_{1}$, the contour-ordered Green function is
\begin{eqnarray}
G_{\underbrace{2...2}_{\ell}\underbrace{1...1}_{n{-}\ell}} (x_{1},...
x_{n})&&\label{G2211x}\\
&&\hskip-2cm =({-}i)^{n{-}1}\sum_{a,b}C_{ab}\langle
a|\widetilde{T}\bigl[\phi(x_{1})...\phi(x_{\ell})\bigr]|b\rangle\nonumber\\
&&\times\langle b|T\big[\phi(x_{\ell{+}1})...\phi(x_{n})\big]|a\rangle\nonumber
,\end{eqnarray} where $C_{ab}$ is the same as in Eq. (\ref{cab}).
Because  $\sigma$ is in the range $0\le \sigma\le \beta$,
$C_{ab}$ will decrease as either $E_{a}$ or $E_{b}$ become large.  Eq.
(\ref{G2211x}) is invariant under permutations among the
first
$\ell$ coordinates and invariant under permutations among the last $n-\ell$
coordinates.  Any permutation that interchanges one of the first $\ell$
coordinates with one of the last $n-\ell$ will give a new function.
 
\subsection{Fourier transforms}

Invariance under time translation guarantees that the Fourier transforms will have
the form

\begin{eqnarray}
\int\prod_{j=1}^{n}\Big[d^{4}x_{j}
e^{ip_{j}\cdot
x_{j}}\Big]G_{\alpha_{1}\alpha_{2}...\alpha_{n}}(x_{1},x_{2},...x_{n})\\
=2\pi\delta\Big[\sum_{s=1}^{n}p_{s}^{0}\Big]
G_{\alpha_{1}\alpha_{2}...\alpha_{n}}(p_{1},p_{2},...p_{n})
\end{eqnarray}
As before, $G(p)$ is proportional to a Dirac delta function that conserves
three-momentum. The constraint of energy conservation,
\begin{equation}p_{1}^{0}{+}p_{2}^{0}{+}p_{3}^{0}{+}...{+}p_{n}^{0}=0,
\label{constraint}\end{equation}
will be very important in what follows. 

\subsubsection{Fourier transform of $G_{11\dots 1}$.}
The Fourier transform of Eq.
(\ref{1111}) can  be written as a sum over all permutations:
\begin{eqnarray}
&&G_{1...1}(p_{1},...p_{n}) \label{1111p}\\
&&=\sum_{a}\sum_{[\,\text{perm}\,]}P_{a}
\langle
a|\phi_{[1]}{\cal G}^{R}_{[1]}\phi_{[2]}{\cal
G}^{R}_{[12]}...
\phi_{[n]}|a\rangle,
\label{G1111}\nonumber\end{eqnarray}
where  $[\;]$ is a permutation   that sends the
ordered set
$\{1,2,... ,n\}$ into the ordered set $\{[1], [2],...,[n]\}$.
The notation is the same as Eqs. (\ref{Pa}),  (\ref{space}),
and(\ref{ret}). The retarded resolvents  depend on the sums
of various $p_{j}^{0}$: 
\begin{eqnarray}
&&{\cal G}_{ij}^{R}={1\over
E_{a}{+}p_{i}^{0}{+}p_{j}^{0}{-}H{+}i\epsilon}\nonumber\\
&& {\cal
G}_{ijk}^{R}={1\over
E_{a}{+}p_{i}^{0}{+}p_{j}^{0}{+}p_{k}^{0}{-}H{+}i\epsilon}\nonumber\\ 
&&{\cal
G}_{ijk\ell}^{R}={1\over
E_{a}{+}p_{i}^{0}{+}p_{j}^{0}{+}p_{k}^{0}{+}p_{\ell}^{0}{-}H{+}i\epsilon}
\nonumber.
\nonumber\end{eqnarray}

\paragraph*{Zero-temperature limit.} At $T=0$ only the vacuum state $|0\rangle$
contributes to Eq. (\ref{1111p}). The n-point function is
\begin{displaymath}
G_{1...1}(p_{1},...p_{n})\to\sum_{[\,\text{perm}\,]}
\langle
0|\phi_{[1]}\,{\cal G}^{R}_{[1]}\phi_{[2]}{\cal
G}^{R}_{[12]}...
\phi_{[n]}|0\rangle,
\end{displaymath}
but with $E_{a}= 0$ in the resolvents:
\begin{displaymath}
{\cal G}^{R}_{z}\to{1\over z-H+i\epsilon}.\end{displaymath}
Since the spectrum of the Hamiltonian is positive,
the vacuum resolvent 
produces no branch cuts when  $z$ is negative. 
For $2\to 2$ scattering, two energies $p_{1}^{0}$ and $p_{2}^{0}$ are positive
real; two energies $p_{3}^{0}$ and $p_{4}^{0}$ are negative real.
Thus at zero temperature there is no problem in analytically extending the
4-point function  off the real axis to allow
 $p_{1}^{0}$ and $p_{2}^{0}$ to lie in the first quadrant and 
$p_{3}^{0}$ and $p_{4}^{0}$ to lie in the third quadrant. 

 \subsubsection{Fourier transform of $G_{22...2}$}

The Green function in which all fields are anti-time-ordered
is given in Eq. (\ref{G2222x}) and has the Fourier transform
\begin{eqnarray}
&&G_{2...2}(p_{1},...p_{n})\label{2222p} \\
&&=(-1)^{n{-}1}\sum_{a}\sum_{[\,\text{perm}\,]}P_{a}
\langle
a|\phi_{[1]}{\cal G}^{A}_{[1]}\phi_{[2]}{\cal
G}^{A}_{[12]}...
\phi_{[n]}|a\rangle,
\label{G2222}\nonumber\end{eqnarray}
The denominators of the advanced resolvents have a negative imaginary part,
$-i\epsilon$ as in Eq. (\ref{adv}), whereas the retarded have a positive
imaginary part,
$+i\epsilon$.

\subsubsection{Fourier transform of $G_{2...21...1}$}

In Eq. (\ref{G2211x}) the times $t_{1},t_{2},...t_{\ell}$ are independent
of the times $t_{\ell{+}1},...t_{n}$ and consequently the Fourier
transform of $G_{2...21...1}$ contains
the Dirac delta function
\begin{equation}
 i2\pi
\delta\Big[E_{a}\!+\!\sum_{j=1}^{\ell}p_{j}^{0}\!-\!E_{b}\Big].
\label{delta}
\end{equation}
This can be written as the difference between an advanced resolvent and a
retarded resolvent:
\begin{displaymath}
\langle b|\,{\cal G}^{A}_{12...\ell}-{\cal G}^{R}_{12...\ell}\,|b\rangle.
\end{displaymath}
The  Fourier transform of Eq. (\ref{G2211x}) is
\begin{widetext}
\begin{eqnarray}
&&G_{\underbrace{2...2}_{\ell}\underbrace{1...1}_{n-\ell}}(p_{1},p_{2},...p_{n})=
(\!-1)^{\ell}e^{-\sigma\sum_{1}^{\ell}p_{j}^{0}}
\sum_{a} P_{a}\sum_{[\,\text{perm}\,']}\label{2211p}\\
&&\hskip2cm \langle a|\big(\phi_{[1]}
{\cal G}^{A}_{[1]}\phi_{[2]}{\cal G}^{A}_{[12]}\dots
\phi_{[\ell]}\big)
\big({\cal G}_{12...\ell}^{A}-{\cal G}_{12...\ell}^{R}\big)
\big(\phi_{[\ell+1]}\,{\cal
G}^{R}_{12...\ell[\ell{+}1]}\phi_{[\ell+2]}
\,{\cal G}^{R}_{12...\ell[\ell{+}1,\ell{+}2]}\dots
\phi_{[n]}\big)|a\rangle,\nonumber
\end{eqnarray}
where $[\,\text{perm}\,']$ are permutations that do not mix the set
$\{1,2,...\ell\}$ with the set $\{\ell{+}1,...n\}$.
A useful relation is
\begin{eqnarray}
&&e^{-\beta\sum_{1}^{\ell}p_{j}^{0}}\;G_{\underbrace{2...2}_{\ell}\underbrace{1...1}_{n-\ell}}(p_{1},p_{2},...p_{n})=
(\!-1)^{\ell}e^{-\sigma\sum_{1}^{\ell}p_{j}^{0}}
\sum_{a} P_{a}\sum_{[\,\text{perm}\,']}\label{2211palt}\\
&&\hskip2cm \langle a|
\big(\phi_{[\ell+1]}\,{\cal
G}^{R}_{[\ell{+}1]}\phi_{[\ell+2]}
\,{\cal G}^{R}_{[\ell{+}1,\ell{+}2]}\dots
\phi_{[n]}\big)
\big({\cal G}_{\ell{+}1...n}^{A}-{\cal G}_{\ell{+}1...n}^{R}\big)\big(\phi_{[1]}
{\cal G}^{A}_{[1]\ell{+}1...n}\phi_{[2]}{\cal G}^{A}_{[12]\ell{+}1...n}\dots
\phi_{[\ell]}\big)|a\rangle,\nonumber
\end{eqnarray}
\end{widetext}
which follows by using Eq. (\ref{delta}) and interchanging the states
$|a\rangle\leftrightarrow |b\rangle$.
It is important to note that in Eq. (\ref{2211p}) all the advanced resolvents
${\cal G}^{A}$ are associated with external lines of type 2. For example,
all the functions
$G_{2211}$, $G_{1122}$, $G_{1212}$, $G_{2121}$, $G_{1221}$, and $G_{2112}$ have
two ${\cal G}^{A}$ resolvents on the left and two ${\cal G}^{R}$
resolvents on the right. Eq. (\ref{2211palt}) has the resolvent pattern
reversed. 

\subsection{Impossibility of analytic extensions}

\subsubsection{n=3 Example}
This simplest way to see the impossibility of analytic extension is to examine a
specific case, viz 
$n=3$. 
The energies for all  the three-point functions are
constrained by
\begin{equation}
p_{1}^{0}+p_{2}^{0}+p_{3}^{0}=0.\label{3energies}\end{equation} 

\paragraph*{$G_{111}$ has no analytic extension.}
The time-ordered function $G_{111}(x)$
contains $3!{=}6$ different time orderings and this leads to the Fourier
transform being the sum of six different  matrix elements of the type in Eq.
(\ref{1111p}):
\begin{eqnarray}
&&G_{111}(p_{1},p_{2},p_{3})\label{111n=3}\\
&&\hskip1.3cm=\sum P_{a}\langle
a|\phi_{1}{\cal G}^{R}_{1}\phi_{2}{\cal
G}^{R}_{12}\phi_{3}+
\phi_{1}{\cal G}^{R}_{1}\phi_{3}{\cal
G}^{R}_{13}\phi_{2}\nonumber\cr
&&\hskip2.8cm+\phi_{2}{\cal G}^{R}_{2}\phi_{1}{\cal
G}^{R}_{12}\phi_{3}
+\phi_{2}{\cal G}^{R}_{2}\phi_{3}{\cal
G}^{R}_{23}\phi_{1}\nonumber\\
&&\hskip2.8cm+\phi_{3}{\cal G}^{R}_{3}\phi_{1}{\cal
G}^{R}_{13}\phi_{2}
+\phi_{3}{\cal G}^{R}_{3}\phi_{2}{\cal
G}^{R}_{23}\phi_{1}|a\rangle.\nonumber
\end{eqnarray}
The one-particle resolvent ${\cal G}_{1}^{R}$ can be analytically
extended to
$\text{Im}(p_{1}^{0})>0$\,;
${\cal G}_{2}^{R}$ can be analytically extended to
$\text{Im}(p_{2}^{0})>0$\,; and ${\cal G}_{3}^{R}$ can be analytically
extended to $\text{Im}(p_{3}^{0})>0$. These three requirements are
incompatible with the constraint (\ref{3energies}).

\paragraph*{$G_{211}$ has no analytic extension.}  By Eq.
(\ref{2211p})
\begin{eqnarray}
&&e^{\sigma p_{1}^{0}}G_{211}(p_{1},p_{2},p_{3})
\label{211n=3}\\
&&\hskip0.8cm =\sum_{a}P_{a}\langle a|\phi_{1}\big({\cal
G}_{1}^{R}{-}{\cal G}_{1}^{A}\big)\big(\phi_{2}{\cal
G}_{12}^{R}\phi_{3}+\phi_{3}{\cal G}_{13}^{R}
\phi_{2}\big)|a\rangle.\nonumber
\end{eqnarray}
The difference ${\cal G}_{1}^{R}-{\cal
G}_{1}^{A}=-i2\pi\delta[E_{a}{+}p_{1}^{0}{-}H]$ forces $p_{1}^{0}$ to be real.
For $p_{1}^{0}$ real, an analytic extension of ${\cal G}_{12}^{R}$ would
require
$\text{Im}(p_{2}^{0})>0$\,; analytic extension of ${\cal G}_{13}^{R}$ 
would require
$\text{Im}(p_{3}^{0})>0$. These two conditions are incompatible with
Eq. (\ref{3energies}).

\paragraph*{$G_{211}$ is not a discontinuity.} Although $G_{211}$ can only 
be defined for real energies, it has the appearance of a discontinuity.
Define two functions $E$ and $F$ so that
\begin{displaymath}
e^{\sigma
p_{1}^{0}}G_{211}(p_{1},p_{2},p_{3})=E(p_{1},p_{2},p_{3})-F(p_{1},p_{2},p_{3})
\end{displaymath}
Although $E$ and $F$ are defined only for real energies, it might be possible to
extend them analytically into two different domains. Their difference would then
be the discontinuity across the branch cut in $p_{1}^{0}$. The explicit functions
are
\begin{eqnarray}
&&E(p_{1},p_{2},p_{3})\nonumber\\
&&\hskip1cm =\sum_{a}P_{a}\langle
a|\phi_{1}{\cal G}_{1}^{R}\phi_{2}{\cal
G}_{12}^{R}\phi_{3}+\phi_{1}{\cal G}_{1}^{R}\phi_{3}{\cal G}_{13}^{R}
\phi_{2}\big)|a\rangle;\nonumber\\
&&F(p_{1},p_{2},p_{3})\nonumber\\
&&\hskip1cm =\sum_{a}P_{a}\langle
a|\phi_{1}{\cal G}_{1}^{A}\phi_{2}{\cal
G}_{12}^{R}\phi_{3}+\phi_{1}{\cal G}_{1}^{A}\phi_{3}{\cal G}_{13}^{R}
\phi_{2}\big)|a\rangle.\nonumber
\end{eqnarray}
$E(p)$ can be analytically extended to  the region in which
$\text{Im}(p_{1}^{0})>0$,
$\text{Im}(p_{1}^{0}{+}p_{2}^{0})>0$, and
$\text{Im}(p_{1}^{0}{+}p_{3}^{0})>0$. The latter two conditions imply
$\text{Im}(p_{3}^{0})<0$
 and $\text{Im}(p_{2}^{0})<0$, which are consistent with Eq.
(\ref{3energies}). However, $F(p)$ cannot be analytically extended
because
${\cal G}_{1}^{A}$ requires $\text{Im}(p_{1}^{0})<0$\,;
${\cal G}_{12}^{R}$
requires $\text{Im}(p_{1}^{0}{+}p_{2}^{0})>0$  [which makes
$\text{Im}(p_{3}^{0})<0$\,]; and ${\cal G}_{13}^{R}$
requires $\text{Im}(p_{1}^{0}{+}p_{3}^{0})>0$  [which makes
$\text{Im}(p_{2}^{0})<0$\,].  It is not possible to have
$\text{Im}(p_{1}^{0})$, $\text{Im}(p_{2}^{0})$, and $\text{Im}(p_{3}^{0})$
all negative because of (\ref{3energies}). 

\paragraph*{Two more examples.} For later purposes it is helpful to display two
more of the mixed functions that can be obtained from Eq. (\ref{2211p}):
\begin{eqnarray}
&&e^{\sigma p_{2}^{0}}G_{121}(p_{1},p_{2},p_{3}) \label{121n=3}\\
&&\hskip0.8cm=\sum_{a}P_{a}
\langle a|\phi_{2}\big({\cal G}_{2}^{R}-{\cal G}_{2}^{A}\big)
\big(\phi_{1}{\cal G}_{12}^{R}\phi_{3}
+\phi_{3}{\cal G}_{23}^{R}\phi_{1}|a\rangle\nonumber\\
&&e^{\sigma p_{3}^{0}}G_{112}(p_{1},p_{2},p_{3}) \label{112n=3}\\
&&\hskip0.8cm =\sum_{a}P_{a}
\langle a|\phi_{3}\big({\cal G}_{3}^{R}-{\cal G}_{3}^{A}\big)
\big(\phi_{1}{\cal G}_{13}^{R}\phi_{2}
+\phi_{2}{\cal G}_{23}^{R}\phi_{1}|a\rangle\nonumber\\
&&e^{\sigma(p_{2}^{0}{+}p_{3}^{0})}G_{122}(p_{1},p_{2},p_{3})\\
&&\hskip0.8cm =\sum_{a}P_{a}\langle a|
\big(\phi_{2}{\cal G}_{2}^{A}\phi_{3}{+}\phi_{3}{\cal G}_{3}^{A}\phi_{2}\big)
\big({\cal G}_{23}^{A}{-}{\cal G}_{23}^{R}\big)\phi_{1}|a\rangle.\nonumber
\end{eqnarray}
These will be used in Sec. IVA.

\subsubsection{Arbitrary n}

The observations for $n=3$ apply generally.
None of the functions $G_{\alpha_{1}...\alpha_{n}}(p_{1},...p_{n})$
displayed in Eqs. (\ref{1111p}), (\ref{2222p}), or (\ref{2211p})
can be extended analytically to  complex energies.

\section{Analyticity of Retarded and advanced \lowercase{n}-point functions}

 It is known 
that certain simple linear combinations of the 
$G_{\alpha_{1}...\alpha_{n}}(p_{1},...p_{n})$ can be extended analytically to
complex energies. These will be reviewed here.  Section V will show that there
are no other linear combinations that can be analytically extended. 

\subsection{$\mathbf{n=3}$ Example.} For the three-point functions, the
constraint (\ref{3energies}) requires that any analytic extension must be to a
domain in which at least one energy has a negative imaginary part and at least
one has a positive imaginary part. However the signs of
$\text{Im}(p_{j}^{0})$ must all be consistent with the retarded or advanced
nature of the resolvents. For example, if $\text{Im}(p_{1}^{0})>0$ the
corresponding resolvent must be retarded: ${\cal G}_{1}^{A}$. The
constraint (\ref{3energies}) implies that
$\text{Im}(p_{2}^{0}{+}p_{3}^{0})<0$ and the complementary resolvent must
be advanced: ${\cal G}_{23}^{A}$. Choosing $\text{Im}(p_{2}^{0})$  and 
$\text{Im}(p_{3}^{0})$ both negative  determines all the resolvents as
follows: 
\begin{equation}
\begin{array}{ccccc}
\text{Im}(p_{1}^{0})>0 &\rightarrow &\text{Im}(p_{2}^{0}{+}p_{3}^{0})<0
&\rightarrow &{\cal G}_{1}^{R}\,\text{and}\,{\cal G}_{23}^{A}\\
\text{Im}(p_{2}^{0})<0 &\rightarrow &\text{Im}(p_{1}^{0}{+}p_{3}^{0})>0
&\rightarrow & {\cal G}_{2}^{A}\,\text{and}\, {\cal G}_{13}^{R}\\
\text{Im}(p_{3}^{0})<0 &\rightarrow &\text{Im}(p_{1}^{0}{+}p_{2}^{0})>0
&\rightarrow &{\cal G}_{3}^{A}\,\text{and}\,{\cal G}_{12}^{R}.
\end{array}\label{3chart}
\end{equation}
The three-point function that is analytic in this domain will be called
$R(\buildrel\circ\over{p_{1}},p_{2},p_{3})$. The circumflex over
$p_{1}$  indicates that only $p_{1}^{0}$ can have a positive imaginary
part. All the resolvents are determined in Eq. (\ref{3chart}) and give
\begin{eqnarray}
&&R(\buildrel\circ\over{p_{1}},p_{2},p_{3})\label{3ret}\\
&&\hskip1.1cm=\sum P_{a}\langle
a|\phi_{1}{\cal G}^{R}_{1}\phi_{2}{\cal
G}^{R}_{12}\phi_{3}+
\phi_{1}{\cal G}^{R}_{1}\phi_{3}{\cal
G}^{R}_{13}\phi_{2}\nonumber\cr
&&\hskip2.6cm+\phi_{2}{\cal G}^{A}_{2}\phi_{1}{\cal
G}^{R}_{12}\phi_{3}
+\phi_{2}{\cal G}^{A}_{2}\phi_{3}{\cal
G}^{A}_{23}\phi_{1}\nonumber\\
&&\hskip2.6cm+\phi_{3}{\cal G}^{A}_{3}\phi_{1}{\cal
G}^{R}_{13}\phi_{2}
+\phi_{3}{\cal G}^{R}_{A}\phi_{2}{\cal
G}^{A}_{23}\phi_{1}|a\rangle.\nonumber
\end{eqnarray}
This can be expressed as a linear combination of the
$G_{\alpha_{1}\alpha_{2}\alpha_{3}}$ as follows.  To change
$G_{111}(p_{1},p_{2},p_{3})$  in Eq. (\ref{111n=3}) into (\ref{3ret}), the 
first step is to change 
 ${\cal G}_{2}^{R}$ into  ${\cal G}_{2}^{A}$.
This can be done by subtracting  Eq. (\ref{121n=3}) from Eq. (\ref{111n=3}):
\begin{eqnarray}
&&G_{111}(p_{1},p_{2},p_{3})
-e^{\sigma p_{2}^{0}}G_{121}(p_{1},p_{2},p_{3})\nonumber\\
&&\hskip1.3cm=\sum P_{a}\langle
a|\phi_{1}{\cal G}^{R}_{1}\phi_{2}{\cal
G}^{R}_{12}\phi_{3}+
\phi_{1}{\cal G}^{R}_{1}\phi_{3}{\cal
G}^{R}_{13}\phi_{2}\nonumber\cr
&&\hskip2.8cm+\phi_{2}{\cal G}^{A}_{2}\phi_{1}{\cal
G}^{R}_{12}\phi_{3}
+\phi_{2}{\cal G}^{A}_{2}\phi_{3}{\cal
G}^{R}_{23}\phi_{1}\nonumber\\
&&\hskip2.8cm+\phi_{3}{\cal G}^{R}_{3}\phi_{1}{\cal
G}^{R}_{13}\phi_{2}
+\phi_{3}{\cal G}^{R}_{3}\phi_{2}{\cal
G}^{R}_{23}\phi_{1}|a\rangle.\nonumber
\end{eqnarray}
Next subtract $e^{\sigma p_{3}^{0}}G_{112}(p_{1},p_{2},p_{3})$ in order to change
${\cal G}_{3}^{R}$ into ${\cal G}_{3}^{A}$. Then 
add $e^{\sigma(p_{2}^{0}{+}p_{3}^{0})}G_{122}$
to change ${\cal G}_{23}^{R}$ into ${\cal G}_{23}^{A}$. This gives the retarded
function as a linear combination of the $G_{\alpha_{1}\alpha_{2}\alpha_{3}}$:
\begin{eqnarray}
R(\buildrel\circ\over{p_{1}},p_{2},p_{3})&=&
G_{111}(p_{1},p_{2},p_{3})
-e^{\sigma p_{2}^{0}}G_{121}(p_{1},p_{2},p_{3})\nonumber\\
&&-e^{\sigma p_{3}^{0}}G_{112}(p_{1},p_{2},p_{3})\nonumber\\
&&+e^{\sigma( p_{2}^{0}{+}p_{3}^{0})}G_{122}(p_{1},p_{2},p_{3})
\label{3ret'}
\end{eqnarray}
It is worth emphasizing that none of the $G_{\alpha_{1}\alpha_{2}\alpha_{3}}$
on the right hand side can be extended analytically to complex energies. However
the combination on the right hand side can be extended analytically  to the region
given in Eq. (\ref{3chart}).

\subsection{Retarded n-point functions and their analytic extension}

Eq. (\ref{3ret'}) is a special case of known results \cite{Chris1,Chris2}.
  The n-point retarded function can be
expressed as the following linear combination 
\begin{eqnarray}
&&R(\buildrel\circ\over{p_{1}},p_{2},...p_{n})
=\sum_{\alpha_{2}=1}^{2}\sum_{\alpha_{3}=1}^{2}...\sum_{\alpha_{n}=1}^{2}\\
&&\hskip2cm
\times C^{R}_{\alpha_{2}...\alpha_{n}}
G_{1\alpha_{2}\alpha_{3}...\alpha_{n}}(p_{1},p_{2},...p_{n}),
\nonumber\end{eqnarray}
where the coefficients are
\begin{displaymath}
C^{R}_{\alpha_{2}...\alpha_{n}}=\prod_{j=2}^{n}
\Big[-e^{\sigma p_{j}^{0}}\Big]^{\alpha_{j}-1}.
\end{displaymath}
The  Fourier transform has the structure
\begin{equation}
R(\buildrel \circ\over{p_{1}}, p_{2},...p_{n})
=\sum_{S=1}^{n!}R_{S}.\label{nret}
\end{equation}
Each $R_{S}$ is a thermal average of a particular permutation of the fields
$\phi_{1}, \phi_{2}, ...\phi_{n}$ of the form
\begin{displaymath}
R_{S}
=\sum_{a}P_{a}\langle a|\phi_{i}{\cal G}^{A}\phi_{j}{\cal G}^{A}
...\phi_{k}{\cal G}^{A}\buildrel\downarrow\over\phi_{1}{\cal G}^{R}\phi_{l}
{\cal G}^{R}
...{\cal G}^{R}\phi_{m}|a\rangle.
\end{displaymath}
Wherever $\phi_{1}$ occurs in the permutation, all the resolvents to the left
are advanced ${\cal G}^{A}$ and all the resolvents to the right are retarded ${\cal
G}^{R}$.   The energy dependence of each resolvents is
cumulative from left to right. [See Eq. (\ref{3ret}).]

\paragraph*{Analytic extension of the n-point retarded function.}
All $n!$  terms  in Eq.
(\ref{nret}) can be extended analytically  to the complex energy domain
 such that
$p_{1}^{0}$ has a positive imaginary part and  all the
other energies
$p_{2}^{0}, p_{3}^{0},...p_{n}^{0}$  have negative imaginary
parts.  To verify this statement requires two steps.

 (i) The first
step is to confirm that   the signs of the imaginary parts of all the
subenergies are determined. Any partial sum
$p_{c}^{0}{+}p_{d}^{0}{+}...p_{r}^{0}$ may or may not include
$p_{1}^{0}$. If it does not include $p_{1}^{0}$ then the
imaginary part of the partial sum is automatically negative since
$\text{Im}(p_{1}^{0})>0$. If the partial sum does include $p_{1}^{0}$, then
the imaginary part of the partial sum is automatically positive because 
the omitted $p_{j}^{0}$ has a negative imaginary part. 

 (ii) The second step is to confirm that every $R_{S}$ can be analytically extended
to this domain.  In $R_{S}$ the advanced resolvents
${\cal G}^{A}$ all depend on partial sums of the energies
$p_{i}^{0}{+}p_{j}^{0}{+}...p_{m}^{0}$ that do not contain
$p_{1}^{0}$ and therefore these advanced resolvents can be
extended analytically to the lower half-plane as described in
(i). In $R_{S}$ the retarded resolvents
${\cal G}^{R}$ all depend on partial sums of the energies
$p_{i}^{0}{+}p_{j}^{0}{+}...p_{m}^{0}$ that do  contain
$p_{1}^{0}$ and therefore these advanced resolvents can be
extended analytically to the upper half-plane as described in
(i).

\subsection{Advanced n-point functions and their analytic extension}

 In a parallel fashion the advanced n-point
functions are also linear combinations of the
$G_{1\alpha_{2}\alpha_{3}...\alpha_{n}}$ as follows:
\begin{eqnarray}
&&A(p_{1},\buildrel\circ\over{p_{2}},\buildrel\circ\over{p_{3}},
...\buildrel\circ\over{p_{n}})
=\sum_{\alpha_{2}=1}^{2}\sum_{\alpha_{3}=1}^{2}...\sum_{\alpha_{n}=1}^{2}
\label{nadv}\\
&&\hskip2cm
\times C^{A}_{\alpha_{2}...\alpha_{n}}
G_{1\alpha_{2}\alpha_{3}...\alpha_{n}}(p_{1},p_{2},...p_{n}),
\nonumber\end{eqnarray}
where now the coefficients are
\begin{displaymath}
C^{A}_{\alpha_{2}...\alpha_{n}}=\prod_{j=2}^{n}
\Big[-e^{(\sigma-\beta) p_{j}^{0}}\Big]^{\alpha_{j}-1}.
\end{displaymath}
Eq. (\ref{2211palt}) is useful here. 
The Fourier transform of the
advanced function is a sum of thermal averages of 
$n!$ terms:
\begin{equation}
A(p_{1},
\buildrel \circ\over{p_{2}},...\buildrel \circ\over{p_{n}})
=\sum_{S=1}^{n!}A_{S}.\end{equation}
Each $A_{S}$ is a particular permutation of the $\phi$'s:
\begin{displaymath}
A_{S}=\sum_{a}P_{a}
\langle a|\phi_{i}{\cal G}^{R}\phi_{j}{\cal G}^{R}
...\phi_{k}{\cal G}^{R}\buildrel\downarrow\over\phi_{1}{\cal
G}^{A}\phi_{l} {\cal G}^{A} ...{\cal
G}^{A}\phi_{m}|a\rangle.
\end{displaymath}
 To the left of $\phi_{1}$ all
the resolvents are retarded and to the right of $\phi_{1}$ all the
resolvents are advanced. The energy dependence of each resolvents is
cumulative from left to right. 

\paragraph*{Analytic extension of the n-point advanced function.}
All $n!$  terms  in Eq.
(\ref{nret}) can be extended analytically  to the complex energy domain
 such that
$p_{1}^{0}$ has a negative imaginary part and  all the
other energies
$p_{2}^{0}, p_{3}^{0},...p_{n}^{0}$  have positive imaginary
parts.  The proof is similar to the retarded case. 

\section{Failure of other combinations to have analytic extensions}

The retarded and advanced functions of Sec. IV show that it is possible to
form linear combinations of the  $G_{\alpha_{1}...\alpha_{n}}(p_{1},...p_{n})$
that can be analytically extended to  certain domains of  complex
energy.  The simplicity of those results may  suggest that other
linear combinations  can be analytically extended to other domains of
complex energy. This section will demonstrate that there are no other such
combinations.  

The key feature is the pattern of resolvents.
In $G_{111...111}$ the resolvents are all retarded as in Eq. (\ref{1111p}).
In $G_{222...222}$ the resolvents are all advanced as in Eq. (\ref{2222p}).
In $G_{2...21...1}$ the resolvents are either a string of advanced followed by
a string of retarded as in Eq. (\ref{2211p}) or a string of retarded followed
by a string of advanced as in Eq. (\ref{2211palt}).  These possibilities
are summarized by
\begin{equation}\begin{array}{cccc}
G_{111...111} &{\cal G}^{R}{\cal G}^{R}{\cal G}^{R}...
{\cal G}^{R}\\
G_{222...222} & {\cal G}^{A}{\cal G}^{A}{\cal G}^{A}...{\cal G}^{A}\\
G_{2...21...1} &{\cal G}^{A}...{\cal G}^{A}{\cal G}^{R}...{\cal G}^{R}
 &\text{or}& {\cal G}^{R}...{\cal G}^{R}{\cal G}^{A}...{\cal G}^{A}.
\end{array}\label{possibilities}
\end{equation}
All $n!$ terms in the retarded n-point functions in Eq. (\ref{nret}) and in the
advanced n-point functions in Eq. (\ref{nadv}) have resolvent patterns of
these types.

The retarded and advanced functions are the exception. No other
thermal n-point function that can be 
analytically extended to some domain of complex energy
can be expressed as a linear combination of the sequences
in Eq. (\ref{possibilities}).
The proof relies on the fact that any n-point function that can be analytically
extended is a  sum of $n!$ terms
\begin{equation}
M(p_{1},p_{2},...p_{n})=\sum_{S=1}^{n!}M_{S}.\label{M}
\end{equation}
Each $M_{S}$ involves one permutation of the
$\phi_{1}$, $\phi_{2}$, ...$\phi_{n}$ and is of the form
\begin{equation}
M_{S}=\sum_{a}P_{a}
\langle a|\phi_{i}{\cal G}_{i}\phi_{j}
{\cal G}_{ij}\phi_{k}{\cal G}_{ijk}...\phi_{r}|a\rangle\end{equation}
The energies in the resolvents, denoted by the subscripts, are cumulative
from left to right. Each resolvent is either retarded or advanced and
that determination is the difficult part of the problem. 
(A simple way to verify this structure is to
extend the imaginary-time n-point function
 from discrete, imaginary energies to
continuous complex energies in various domains as discussed in
 \cite{Tim}.)

The following argument  will show that 
if Eq. (\ref{M}) is analytic when two or more energies have positive
imaginary parts and two or more energies have negative imaginary parts,
then it cannot be a linear combination of the 
basis functions in Sec. III.

Suppose that
$p_{1}^{0}$ and
$p_{2}^{0}$ have positive imaginary parts;  $p_{3}^{0}$ and $p_{4}^{0}$
have negative imaginary parts; and $p_{5}^{0},...p_{n}^{0}$  have
imaginary parts of either sign. This certainly does not  specify
the location of all the  subenergies in the complex plane, but it is all
that will be necessary. Define two sums of $n{-}1$ energies:
\begin{eqnarray}
&&\zeta=\sum_{j\neq 2}p_{j}^{0}\nonumber\\
&&\eta=\sum_{j\neq 4}p_{j}^{0}.
\nonumber\end{eqnarray}
The relation $p_{2}^{0}{+}\zeta=0$ implies that
$\text{Im}(\zeta)<0$\,; the relation $p_{4}^{0}{+}\eta=0$ implies
that $\text{Im}(\eta)>0$. To construct an n-point function that can be
analytically extended, all the resolvents with energy
$\zeta$ must be advanced, viz.
${\cal G}_{\zeta}^{A}$; all the resolvents with energy $\eta$ must be retarded,
viz.
${\cal G}_{\eta}^{R}$. 

Among the $n!$ different  orderings of the $\phi_{j}$,  there will be
terms which begin with  $\phi_{1}$ followed by $\phi_{3}$ and
that end with $\phi_{4}$. When the appropriate resolvents are included,
the    thermal averages have the form
\begin{equation}
\sum_{a}P_{a}
\langle a|\phi_{1}{\cal G}_{1}^{R}\phi_{2}
{\cal G}_{13}^{X}...\phi_{3}...{\cal G}^{R}_{\eta}
\phi_{4}|a\rangle,\label{1st}
\end{equation}
where  $X$ is either retarded or advanced, depending upon the
imaginary part of $p_{1}^{0}{+}p_{3}^{0}$.
Another important class of terms are those that begin with $\phi_{3}$
followed by $\phi_{1}$ and end with $\phi_{2}$. These give  
\begin{equation}
\sum_{a}P_{a}
\langle a|\phi_{3}{\cal G}_{3}^{A}\phi_{1}
{\cal G}_{13}^{X}...\phi_{4}...{\cal G}^{A}_{\zeta}
\phi_{2}|a\rangle.\label{2nd}
\end{equation}
The fact that $p_{1}^{0}$ and $p_{2}^{0}$ have positive imaginary parts
and that $p_{3}^{0}$ and $p_{4}^{0}$ have negative imaginary parts has
determined the resolvents ${\cal G}_{1}^{R}$, ${\cal G}_{\eta}^{R}$, ${\cal
G}_{3}^{A}$, and ${\cal G}_{\zeta}^{A}$.  The value of $X$ in ${\cal
G}_{13}^{X}$ is not yet determined. 

\paragraph*{Case 1.}
If $\text{Im}(p_{1}^{0}{+}p_{3}^{0})>0$ then $X=R$ in
both equations. In Eq. (\ref{2nd}) the resolvents are thus in the order
${\cal G}_{3}^{A}{\cal G}_{13}^{R}...{\cal G}_{\zeta}^{A}$. In other words, at
least one retarded resolvent occurs in a sequence that begins and
ends with advanced resolvents. 
This sequence does not appear in the
list (\ref{possibilities}) and cannot be produced by  linear combinations
of the basis functions in Eqs. (\ref{1111p}), (\ref{2222p}). (\ref{2211p}), or
(\ref{2211palt}).

\paragraph*{Case 2.} If $\text{Im}(p_{1}^{0}{+}p_{2}^{0})<0$ then $X=A$ in
both equations. In Eq. (\ref{1st}) the resolvents are in the order
${\cal G}_{1}^{R}{\cal G}_{13}^{A}...{\cal G}_{\eta}^{R}$. In other words, at
least one advanced resolvent occurs in a sequence that begins and
ends with retarded resolvents. This pattern cannot be constructed out of
linear combinations of the basis functions.
This concludes the proof.

\end{document}